\documentclass{moriond}

\def\be{\begin{equation}}
\def\ee{\end{equation}}
\def\bea{\begin{eqnarray}}
\def\eea{\end{eqnarray}}

\newcommand{\Photo}{\includegraphics[height=35mm]{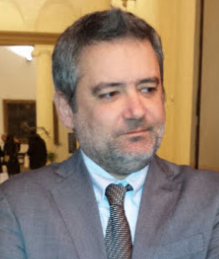}}

\begin{document}
\vspace*{4cm}
\title{54TH RENCONTRES DE MORIOND\\QCD AND HIGH ENERGY INTERACTIONS\\ \vspace{0.5cm}EXPERIMENTAL SUMMARY}

\author{ Vincenzo M. Vagnoni }

\address{Istituto Nazionale di Fisica Nucleare, Sezione di Bologna,\\
via Irnerio 46, 40126 Bologna, Italy\vspace{0.5cm}}

\maketitle\abstracts{
The conference included about 90 talks covering a plethora of different sectors, like Higgs physics, electroweak physics, top physics, BSM physics, soft QCD, jets, PDFs, heavy ions,
heavy flavours, spectroscopy, etc. A few experimental results are here highlighted.}

\section{Introduction}
The Standard Model of particle physics~(SM) has been working beautifully, up to a few hundred GeV so far, yet we know that it must be an effective theory valid up to some energy scale. There are many compelling reasons to believe that the SM in an incomplete theory; we can mention the hierarchy problem, the lack of a dark matter candidate, the unification of gauge couplings, the need for new sources of $C\!P$ violation to explain the dynamical generation of the baryon asymmetry in the universe, the reason for the existence of three generations of quarks and leptons, the origins of mass and CKM hierarchies, etc. The SM has proven to be very solid, well beyond any expectation, but we have powerful tools to go on with the investigations. For example, despite some operational issues, the LHC reached a performance record in 2018, with a peak luminosity steadily at about $2\times10^{34}~\mathrm{cm}^{-2}\mathrm{s}^{-2}$, that is two times larger then the design value.

\section{Higgs physics}
Although it seems with us since ages, we have not to forget that the discovery of the Higgs boson took place only seven years ago. Measurements of Higgs properties with increasing precision are now a formidable tool to look for new-physics manifestations. A framework for cross-section measurements with reduced model dependence targeting Higgs production, namely simplified template cross-sections (STXS), has been developed and deployed. Within this framework, Higgs production is split into the main production modes and further into fine-grain kinematic regions of phase space.

The associated production of $H\to b\bar{b}$ and $W$ or $Z$ bosons decaying into leptons as a function of $p_\mathrm{T}(W/Z)$ with STXS has been measured, using a data sample corresponding to $80~\mathrm{fb}^{-1}$ at 13~TeV~\cite{arXiv:1903.04618}. Cross-sections are used to constrain parameters in an effective Lagrangian framework. Another relevant measurement is that of the Higgs coupling to the top quark using $H\to \gamma\gamma$ decays. A simultaneous fit in seven signal-enriched event categories is performed, and $t\bar{t}H$ production is observed at $4.9\sigma$, using $140~\mathrm{fb}^{-1}$ of data at 13~TeV~\cite{ATLAS-CONF-2019-004}.

\begin{figure}
\begin{minipage}{1\linewidth}
\centerline{\includegraphics[width=0.7\linewidth]{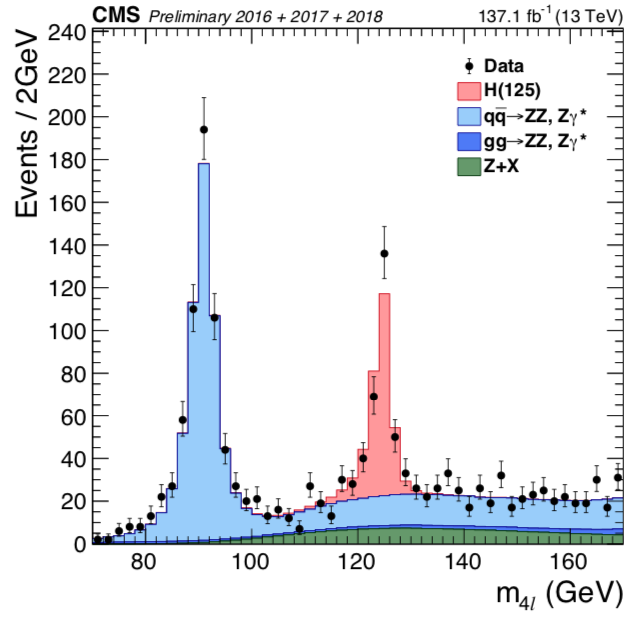}}
\end{minipage}
\caption[]{Distribution of the reconstructed four-lepton invariant mass with 2018 data~\cite{CMS-PAS-HIG-19-001}.}
\label{fig:higgs_mass}
\end{figure}

Several new measurements in the four-lepton final state have been performed. All main production modes have been studied, \emph{i.e.} $ggH$, vector-boson fusion, associated production of gauge bosons and $t\bar{t}H$. There is little sensitivity to $b\bar{b}H$ or $tH$, but these processes have been also considered explicitly. Differential cross-sections have been measured as a function of the Higgs $p_\mathrm{T}$ and $y$, the number of associated jets, and the $p_\mathrm{T}$ of the leading associated jet, using $137~\mathrm{fb}^{-1}$ of data at 13~TeV~\cite{CMS-PAS-HIG-19-001}. The distribution of the reconstructed four-lepton invariant mass with 2018 data is shown in Fig.~\ref{fig:higgs_mass}.

With more and more statistics, searches for Higgs rare decays are becoming increasingly important. Exclusive decays of the Higgs boson to mesons, such as $H \to J/\psi J\psi$ or $\Upsilon\Upsilon$, recently measured using  $37.5~\mathrm{fb}^{-1}$ at 13~TeV~\cite{CMS-PAS-HIG-18-025}, are interesting to study Yukawa couplings to quarks and for beyond-the-SM~(BSM) searches. New physics could affect direct $Hq\bar{q}$ couplings or enter through loops, and modify interference patterns between the various amplitudes. However, SM predictions are very small and uncertain.

We have not yet found evidence for non-SM behaviour of the Higgs particle. But the transition from observation to detailed measurements has only started, and now also in the STXS framework with finer granularity to probe deviations from the SM. Several new Higgs results have been presented at winter conferences, many with partial Run-2 statistics and only a few with full Run-2 statistics. Significant improvements with the full set of analyses and the full Run-2 data set are anticipated.

\section{Top physics}
The LHC is also a powerful top factory. Top quarks are produced predominantly in pairs, with $\sigma(t\bar{t}) \simeq 0.8~\mathrm{nb}$ at 13~TeV. Single-top production via electroweak interactions is rarer, but theoretically very clean. The top quark is the heaviest known fundamental particle by far, with a mass of about 173~GeV. An interesting question is whether the similarity in mass scale to $H$, $W$ and $Z$ is related to electroweak symmetry breaking. Due to the short lifetime, order of $10^{-25}~\mathrm{s}$, the top decays before hadronisation. For this reason it provides a unique opportunity to study a quasi-free quark. Furthermore, spin decorrelation time is much longer, and then one can study spin correlation via decay products~\cite{arXiv:1903.07570}. The top quark decays almost exclusively to $Wb$ in the SM. Events are categorised by the decays of the $W$ boson: all hadronic, lepton plus jets and di-lepton.

The mass of the Higgs at 125 GeV is close to the minimum value that ensures absolute vacuum stability within the SM. Precise top-mass measurements are very relevant, and several different techniques are adopted
\begin{itemize}
\item ideogram method, measured with $35.9~\mathrm{fb}^{-1}$ at 13~TeV~\cite{EPJC 78 (2018) 891};
\item all-jets final state, $35.9~\mathrm{fb}^{-1}$ at 13~TeV~\cite{arXiv:1812.10534};
\item differential cross-section, $35.9~\mathrm{fb}^{-1}$ at 13~TeV~\cite{arXiv:1812.10505};
\item inclusive cross-section, $35.9~\mathrm{fb}^{-1}$ at 13~TeV~\cite{arXiv:1812.10505}.
\end{itemize}

Concerning associated production of $t\bar{t}$ pairs with $W$ and $Z$ bosons, these are rare processes with cross-sections of about $1~\mathrm{pb}$ at 13~TeV. Such measurements are very useful to test QCD predictions, as deviations might indicate the presence of new physics, like vector-like quarks, strongly coupled Higgs bosons, exotic quarks with charge $-4/3$ and anomalous dipole moments of the top quark. Several measurements have been performed recently, \emph{e.g.} by ATLAS with $36.1~\mathrm{fb^{-1}}$ at 13~TeV~\cite{arXiv:1901.03584}, and by CMS with $35.9~\mathrm{fb^{-1}}$ and $77.5~\mathrm{fb^{-1}}$ at 13~TeV~\cite{JHEP 08 (2018) 011,CMS-PAS-TOP-18-009}. Results are in line with NLO predictions. The CMS result on $t\bar{t}Z$ has better precision than NLO, but resummed calculations are now reaching NNLL+NLO accuracy.

An interesting search is that of four-top production. This is a very rare process, with a cross-section of about $10~\mathrm{fb}$ at 13~TeV. Events are characterised by very large jets, b-jet multiplicities and hadronic activity. The latest search by CMS is for same-sign di-lepton and multi-lepton events using $137~\mathrm{fb}^{-1}$ at 13 TeV~\cite{CMS PAS TOP-18-003}. A multivariate analysis yields a significance of $2.6~\sigma$ relative to the background-only hypothesis, and a cross-section of $12.6^{5.8}_{-5.2}~\mathrm{fb}$. The results are used to constrain the top Yukawa coupling with respect to the SM value.

The top quark is also a tool to look directly for BSM physics. In FCNC processes the top quark can couple to a light quark (up or charm) and a neutral boson ($\gamma$, $Z$, $H$, $g$). These processes are forbidden at tree-level in the SM, and not observable with present data unless new physics is at play. Observation of FCNC would be indicative of new physics. A recent measurement by ATLAS in $t\to uH$ and $t \to cH$, using $36.1~\mathrm{fb}^{-1}$ at 13~TeV~\cite{arXiv:1812.11568}, did not find evidence for these processes. Combining the search with other ATLAS searches in di-photon and multi-lepton final states, at 95\% C.L. one gets $BR(t \to uH) < 1.1 \times 10^{-3}$ and $BR(t \to cH) < 1.2 \times 10^{-3}$. There is nothing unexpected in exotic physics with top quarks yet, but the full Run-2 data set has still to be analysed.

To conclude this section, a peculiar measurement of top production that is also worth mentioning is that performed by LHCb, using $1.9~\mathrm{fb}^{-1}$ at 13~TeV~\cite{JHEP 08 (2018) 174}. The forward acceptance of the LHCb detector allows measurements in a phase space inaccessible to ATLAS and CMS. In this analysis, events containing a high-$p_\mathrm{T}$ muon and electron of opposite charge in addition to a high-$p_\mathrm{T}$ jet have been selected. No discrepancy from the SM expectation has been found.

\section{Electroweak physics}
In the SM, three parameters define the electroweak sector, namely U(1) and SU(2) couplings and vacuum expectation value. The electroweak sector of the SM is over-constrained and the strength of global fits can be exploited to predict key observables, such as the $W$ mass and the effective electroweak mixing angle, with a precision exceeding that of direct measurements.

The $W$ mass is sensitive to the Higgs and top masses via radiative corrections. The precise determination of the $W$ mass is of great importance in testing the internal consistency of the SM. The global electroweak fit yields the mass of the $W$ with an uncertainty of 8~MeV, to be compared with the experimental precision of 15~MeV. Therefore there is the need to improve the experimental uncertainty. The only $W$ mass measurement at the LHC to date has been performed by ATLAS: $M_W = 80370 \pm 7\,(\mathrm{stat.}) \pm 11\,(\mathrm{exp. syst.}) \pm 14\,(\mathrm{mod. syst.})$~MeV. The dominating uncertainty stems from theory. This lone ATLAS measurement competes with the Tevatron combination. It is worth reminding that the measurement at the LHC is affected by PDF uncertainties more than at the Tevatron.

In the SM, $ZZ$ production proceeds mainly through quark-antiquark $t$- and $u$-channel scattering diagrams. At higher orders in QCD also $gg$ fusion contributes via box diagrams with quark loops. There has been a recent CMS study of four-lepton production, $pp \to (Z/\gamma^\ast)(Z/\gamma^\ast) \to 4l$, where $l=e~\mathrm{or}~\mu$, using $101~\mathrm{fb}^{-1}$ at 13~TeV~\cite{CMS-PAS-SMP-19-001}. By combining with the 2016 results, the measured cross-section value is found to be consistent with the SM prediction.

\section{BSM searches}
BSM searches are amongst the most important~(and challenging) analyses at the LHC. As first searches, BSM models predict new resonances decaying to a pair of objects, \emph{i.e.} two-body resonances like di-photons, di-leptons, di-bosons, di-jets. When the multiplicity of the final states gets larger, or one goes for exotic signatures like long-lived particles~(LLPs), analyses become increasingly more complex and model-dependent, requiring better understanding of the underlying physics models being tested and of the detector response. Many signatures have been probed already with full Run-2 statistics. First analyses did not reveal evidence for new physics yet, but a lot of phase space has still to be explored. In the following, a few examples of recent analyses presented during the conference will be highlighted.

\begin{figure}
\begin{minipage}{1\linewidth}
\centerline{\includegraphics[width=0.9\linewidth]{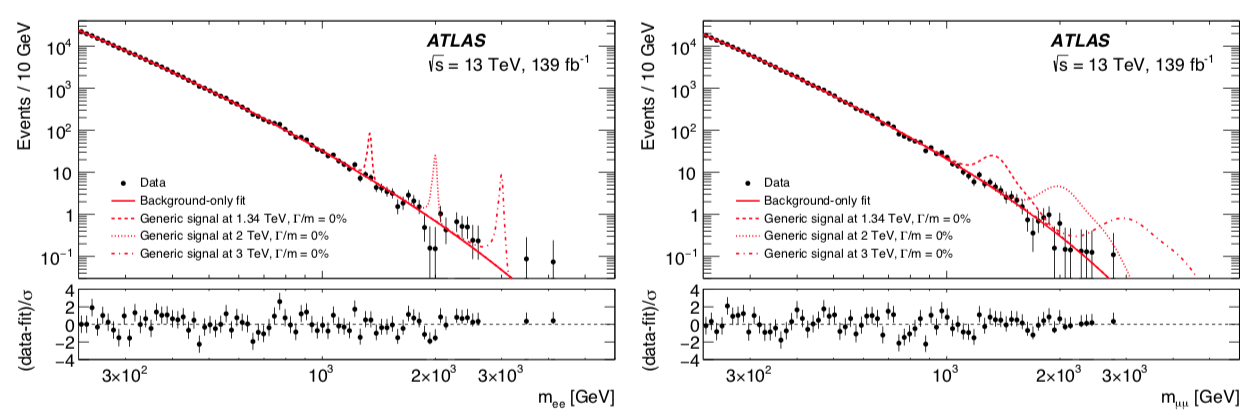}}
\end{minipage}
\caption[]{Distributions of the (left) di-electron and (right) di-muon invariant masses. Zero-width signal shapes, scaled to 20 times the value of the corresponding expected upper limit at 95\% CL on the fiducial cross-section times branching ratio, are superimposed~\cite{arXiv:1903.06248}.}
\label{fig:bsm_mass}
\end{figure}

A recent search for new resonances decaying to electron and muon pairs has been performed by ATLAS using $139~\mathrm{fb}^{-1}$ of data at 13~TeV~\cite{arXiv:1903.06248}. The background fit with a parametric function exhibits an excellent description of the di-lepton spectra up to several TeV. The distributions of the di-electron and di-muon invariant masses are shown in Fig.~\ref{fig:bsm_mass}. No significant excess over the SM expectation is found. Cross-section limits are set for generic resonances with a relative natural width between zero and 10\%.

New heavy particles that decay to partons are predicted in many BSM models. For example, excited quarks are predicted in compositeness models and are a typical benchmark used in many di-jet searches. Partons shower and hadronise, creating collimated jets. BSM phenomena may produce a di-jet signal up to masses that are a significant fraction of the total collision energy. Two recent searches for new resonances in the di-jet invariant mass by ATLAS~(with $139~\mathrm{fb}^{-1}$ at 13 TeV~\cite{ATLAS-CONF-2019-007}) and CMS~(with $77.8~\mathrm{fb}^{-1}$ at 13 TeV~\cite{CMS-PAS-EXO-17-026}) show no significant excess from the SM expectation. Moving to a more complex analysis, with three or more electrons and muons in the final state, one looks for non-resonant excesses in the tails of the sum of lepton $p_\mathrm{T}$ plus missing transverse momentum (with $137~\mathrm{fb}^{-1}$ at 13 TeV~\cite{CMS-PAS-EXO-19-002}). Observed data are consistent with the SM expectation.

While the phase space for an easy discovery is reducing, growing interest is emerging for new-physics searches with unconventional signatures, like emerging jets, heavy charged LLPs, delayed jets, displaced jets, disappearing tracks, displaced muons. A community white paper has been released recently~\cite{arXiv:1903.04497}. At low energy, a search for a spin-0 boson using prompt decays to $\mu^+\mu^-$ has been performed by LHCb, with $3~\mathrm{fb}^{-1}$ of data at 7 and 8 TeV~\cite{JHEP 09 (2018) 147}. The LHCb detector has good sensitivity to light spin-0 particles due to its high-precision spectrometer and its capability of triggering on objects with small transverse momenta. No evidence for a signal is observed and limits are placed.

In conclusion, BSM searches are very challenging as they look for corners and tails of SM physics. The community working in the sector is continuously developing new techniques and adding new models, phase-space regions and correlations with SM backgrounds. Many results are anticipated to come out soon with Run-2 statistics, while preparing for Run 3. The phase space has been narrowed down for some models, but there are many others to study, and the end of the searches is still very far to come.

\section{(Some) hard and soft QCD processes}
Genuine QCD measurements at the LHC are important for the good modelling of hadronic collisions and obviously to test our understanding of QCD: probing PDFs and NLO predictions; studying event topologies in interesting phase space regions, \emph{i.e.} multi-jet production, di-jet decorrelations, and very forward region; studying jet substructure; understanding backgrounds for electroweak analyses, BSM searches, etc.; studying multiple parton interactions, \emph{e.g.} double parton scattering (DPS); etc. In addition, accurate knowledge is crucial for the development of future projects, like ATLAS and CMS phase-2 upgrades. Several recent measurements have been performed by ATLAS and CMS, the latter also using CASTOR to study very forward energy flow and jets.

A measurement of simultaneous Drell-Yan production with four leptons in the final state has been performed using $20.2~\mathrm{fb}^{-1}$ of data at 8 TeV~\cite{Phys. Lett. B 790 (2019) 595}. The process is particularly relevant as a background in the Higgs analysis with four-lepton decays. A simplified model for a DPS cross-section can be written as $\sigma^\mathrm{DPS}_{\mathrm{AB}} \simeq \sigma_\mathrm{A}\sigma_\mathrm{B}/{\sigma_\mathrm{eff}}$, where $\sigma_\mathrm{eff}$ is assumed to be process- and energy-independent. No signal of DPS is observed. The upper limit on the fraction of the DPS contribution to the inclusive four-lepton final state translates into a lower limit of $1.0~\mathrm{mb}$ on the effective cross-section.

Another DPS measurement has been performed by looking for same sign $WW$ production with $77~\mathrm{fb}^{-1}$ of data at 13~TeV~\cite{CMS-PAS-SMP-18-015}. This is an important channel to test DPS predictions. Both hard scatterings lead to the production of a $W$ boson, and particularly interesting is the final state with two same-charge $W$ bosons. The $W$ decay provides a relatively clean signal, with well-understood background processes. In particular, $WZ$ production constitutes the main background. A fit is performed in different flavour-sign categories separately, $\mu^\pm \mu^\pm$ and $e^\pm e^\pm$. This result represents the first experimental evidence of the DPS $WW$ process.

Production of prompt photons in $pp \to \gamma X$ allows pQCD tests with a hard colourless probe. The dominant production mechanism at the LHC is $qg \to q\gamma$. The cross-section is sensitive to the gluon density in the proton. An isolation cut is needed to reduce background from neutral-hadron decays and from fragmentation where the emitted photon is close to a jet. An analysis of $20.2~\mathrm{fb}^{-1}$ and $3.2~\mathrm{fb}^{-1}$ of data at 8 and 13 TeV  has been performed~\cite{arXiv:1901.10075}. The ratio between cross-sections at 13 and 8~TeV is measured as a function of the energy of the photon in different photon pseudorapidity ranges. Predictions using several PDFs agree well with data.

\section{Heavy ions}
Hard probes are one of the pillars to study hot and dense QCD matter created in heavy ion collisions. Perturbative processes take place before the QGP forms, \emph{e.g.} heavy-quark production. On the other hand, heavy quarks decay weakly such that their lifetime is greater than that of the QGP and so they experience the full system evolution. In addition, quarkonium states have binding energies of the order of a few hundred MeV, and interactions with hard gluons in a QGP can overcome this threshold breaking the quarkonium system. The modification of the jet structures while traversing the hot medium is also another relevant example~(jet quenching).

\begin{figure}
\begin{minipage}{1\linewidth}
\centerline{\includegraphics[width=0.7\linewidth]{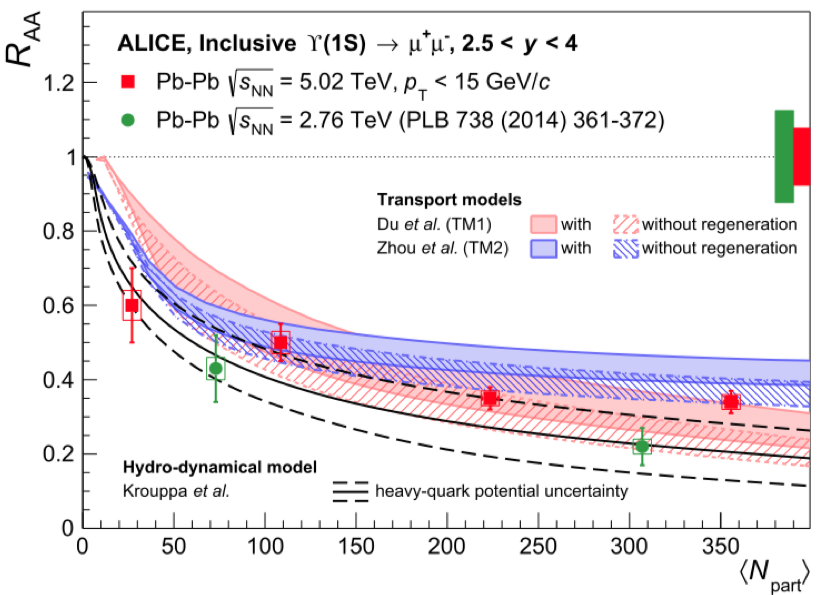}}
\end{minipage}
\caption[]{Nuclear modification factor for $\Upsilon$ production as a function of centrality~\cite{PLB 790 (2019) 89}.}
\label{fig:ALICE_plot}
\end{figure}

The production of $J/\psi$ mesons at high $p_\mathrm{T}$ is strongly suppressed in PbPb with respect to $pp$ collisions. The suppression increases as a function of the centrality of the collision. Also strong $\Upsilon$ suppression is observed (see Fig.~\ref{fig:ALICE_plot}), and higher $\Upsilon$ states are even more suppressed: $R_{AA}(\Upsilon(2S)) / R_{AA}(\Upsilon(1S)) = 0.28 \pm 0.12\,(\mathrm{stat.}) \pm 0.06\,(\mathrm{syst.})$, as measured by ALICE~\cite{PLB 790 (2019) 89}. Also CMS observes a similar spectacular behaviour. The $R_{AA}$ of the $\Upsilon(3S)$ state is measured to be below 0.096 at 95\% C.L.~\cite{Phys. Lett. B 790 (2019) 270}. This is the strongest suppression observed for a quarkonium state in heavy ion collisions to date. By contrast, at low $p\mathrm{T}$, smaller $J/\psi$ suppression at the LHC than at RHIC is found, owing to measurements by ALICE, PHENIX and STAR. New regenerated $J/\psi$ mesons are produced by recombination of charm quarks, with larger regeneration occurring at higher $c\bar{c}$ pair density and higher energy density.

The ratio of $\Lambda_c$ to $D^0$ production has been measured by ALICE in $pp$, $p$Pb, and PbPb collisions~\cite{arXiv:1809.10922}. A similar ratio in $pp$ and $p$Pb collisions is found, whereas enhanced production of $\Lambda_c$ baryons with respect to $D^0$ mesons is observed in PbPb collisions. The measurement is still limited in precision, but appears to be intriguing. Also STAR seems to observe a larger $\Lambda_c$ to $D^0$  ratio in AuAu collisions. In the forward region, as a dedicated heavy-flavour experiment, LHCb has obvious advantages in measuring $b$-hadron decays~(at pPb event multiplicities). Beauty hadrons are cleanly reconstructed in exclusive decay modes~\cite{arXiv:1902.05599}. First measurements in nuclear collisions are made down to very low $p_\mathrm{T}$ values. The $R_{p\mathrm{Pb}}$ suppression pattern seen in $J/\psi$ from $b$ is confirmed. LHCb can also inject gas into the beampipe to measure cross-sections in fixed-target mode, see e.g. antiproton production in pHe~\cite{PRL 121 (2018) 222001} or charm production in pAr and pHe~\cite{arXiv:1810.07907} collisions.

ATLAS measured the flow of muons from heavy flavour decays in $pp$ and PbPb collisions at 2.76~TeV~\cite{Phys. Rev. C 98 (2018) 044905}. The value of $R_{\mathrm{AA}}$ is found to be less than unity, hence observing suppressed production of heavy-flavour muons in PbPb collisions. In particular, for the 10\% most central PbPb events, $R_{\mathrm{AA}}$ is about 0.35. Furthermore, lower values of the elliptic flow are found for particles with heavy quarks than with the lighter quarks. This is a useful information for the analysis of quark interactions in the medium. Charged hadron production in PbPb and XeXe collisions has been studied at 5.02 and 5.44~TeV by ALICE. The production of (most of) light-flavour hadrons in PbPb at 5.02~TeV is described by the thermal model. The differences between protons and the strangeness sector are confirmed. Hydrodynamic properties are studied with spectral shape and azimuthal anisotropy. Hydrodynamics works at low $p_\mathrm{T}$ for central collisions, with the agreement worsening towards peripheral collisions.

In recent years, momentum anisotropies have been measured in $pp$ and $p$A collisions, despite expectations that the volume and lifetime of the medium would be too small. PHENIX observed elliptic and triangular flow patterns in $p$Au, $d$Au and $^3$HeAu collisions~\cite{Nature Physics 15 (2019) 214}. The three initial geometries and two flow patterns provide powerful model discrimination. In the PHENIX analysis, hydrodynamic models, based on the formation of short-lived QGP droplets, provide the best simultaneous description of the measurements.

\section{Heavy flavours}
\begin{figure}
\begin{minipage}{1\linewidth}
\centerline{\includegraphics[width=0.9\linewidth]{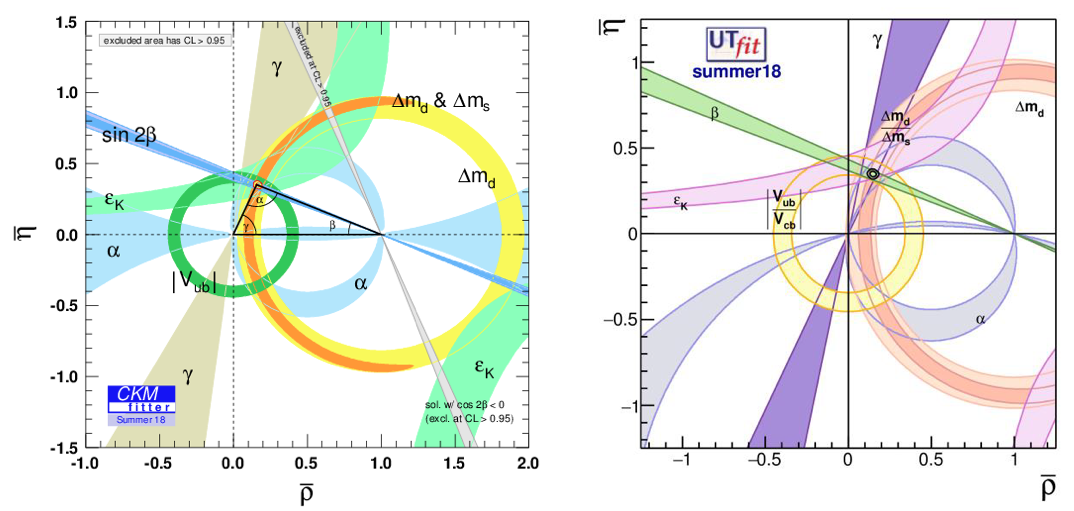}}
\end{minipage}
\caption[]{Results of global CKM fits from (left)~CKMfitter and (right)~UTfit groups.}
\label{fig:CKMfits}
\end{figure}

The present consistency of global CKM fits is displayed in Fig.~\ref{fig:CKMfits}.
Each coloured band defines the allowed region of the apex of the unitarity triangle, according to the measurement of a specific process. Such a consistency represents a tremendous success of the CKM paradigm in the SM: all of the available measurements agree in a highly profound way. In presence of BSM physics affecting the measurements, the various contours would not cross each other into a single point. Hence the quark-flavour sector is generally very well described by the CKM mechanism, and one must look for small discrepancies.

\begin{figure}
\begin{minipage}{1\linewidth}
\centerline{\includegraphics[width=0.9\linewidth]{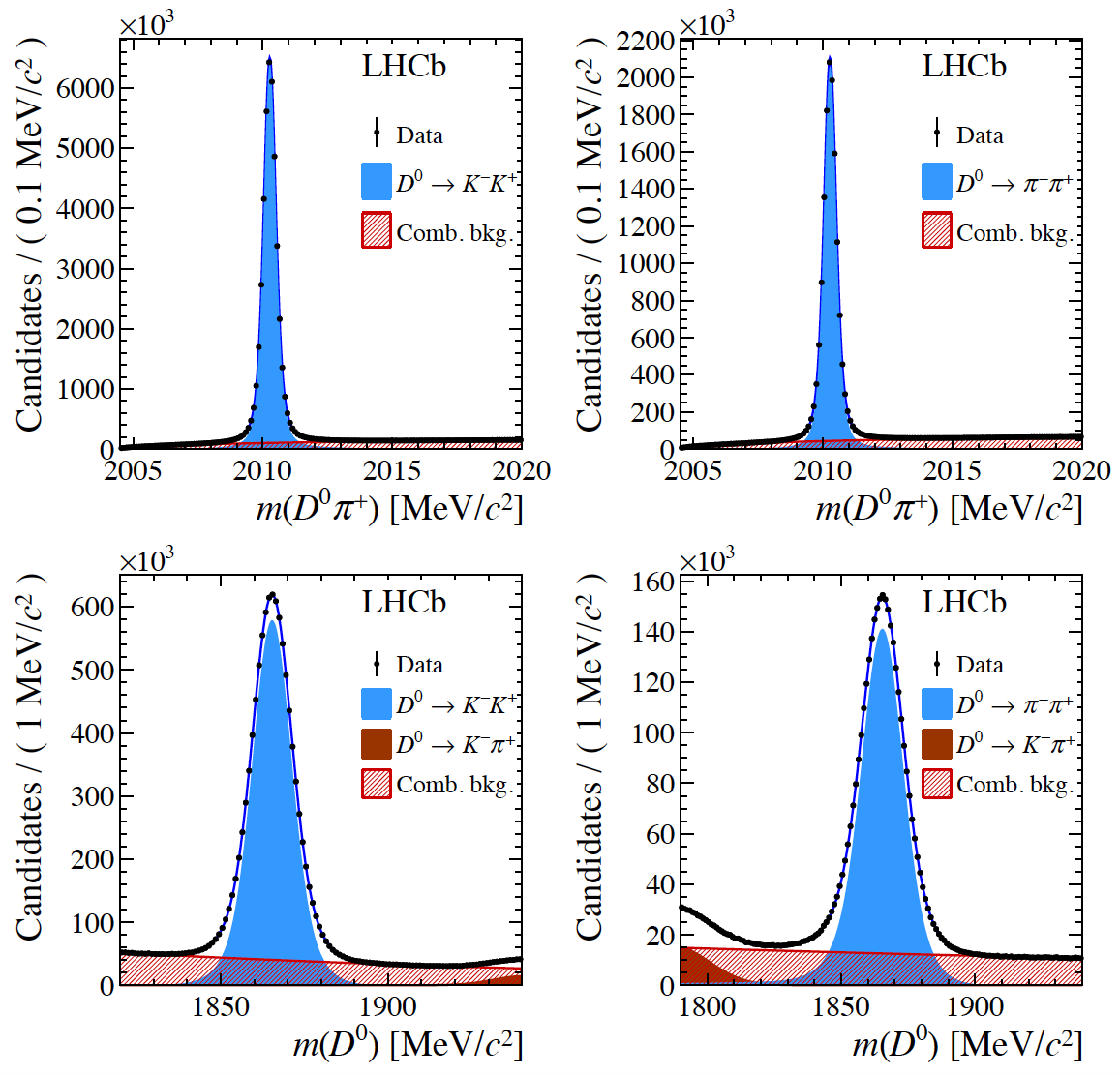}}
\end{minipage}
\caption[]{Mass distributions of selected (top) $\pi^\pm$-tagged and (bottom) $\mu^\pm$-tagged candidates for
(left) $K^+K^-$ and (right) $\pi^+\pi^-$ final states of the $D^0$-meson decays, with fit projections overlaid~\cite{arXiv:1903.08726}.}
\label{fig:lhcb_charm}
\end{figure}

In the charm sector, using $6~\mathrm{fb}^{-1}$ at 13~TeV~\cite{arXiv:1903.08726}, LHCb measured the difference $\Delta A_{C\!P}$ of the $C\!P$-violating asymmetries in the decays $D^0 \to K^-K^+$ and $D^0 \to \pi^-\pi^+$. To perform the measurement, the flavour of the $D^0$ meson is tagged either by using the charge of the pion in $D^{\ast +} \to D^0 \pi^+$ decays, or the charge of the muon in $B \to D^0 \mu \nu X$  decays. The invariant-mass distributions are shown in Fig.~\ref{fig:lhcb_charm}. Run-2 results are well compatible with previous LHCb results and the world average. The combination of Run-1 and Run-2 data gives $\Delta A_{C\!P} = (-15.4 \pm 2.9) \times 10^{-4}$, resulting in the first observation of $C\!P$ violation in the charm sector, with a significance of $5.3\sigma$. The result is roughly compatible with the SM, whose prediction however is way more uncertain than data.

The golden mode $B^0_s \to J/\psi \phi$ is the $B^0_s$ analogue of $B^0 \to J/\psi K^0_\mathrm{S}$, and one measures the interference between $B^0_s$ mixing and decay graphs through the phase-difference $\phi_s$ between the two processes, precisely predicted in the SM to be $\phi_s=-37.4 \pm 0.7$~mrad. This is very small, but can receive sizeable contributions from new physics. New measurements by ATLAS (with $80.5~\mathrm{fb}^{-1}$ at 13~TeV~\cite{ATLAS-CONF-2019-009}) using $B^0_s \to J/\psi \phi$ and LHCb (with $1.9~\mathrm{fb}^{-1}$ at 13~TeV~\cite{LHCb-PAPER-2019-013,arXiv:1903.05530}) using $B^0_s \to J/\psi \phi$ and $B^0_s \to J/\psi \pi\pi$ decays have been performed. The combination of Run-1 and Run-2 data gives $\phi_s = -0.076 \pm 0.034\,(\mathrm{stat}) \pm 0.019\,(\mathrm{syst})$~rad for ATLAS and $\phi_s = -0.040 \pm 0.025\,(\mathrm{stat+syst})$~rad for LHCb. The new world average provided by HFLAV is $\phi_s = -0.0544 \pm 0.0205$~rad. The experimental precision is quickly approaching the sensitivity to observe a nonzero SM value. For this reason, the ATLAS and LHCb analyses with full Run-2 statistics and Run-2 CMS results are now eagerly awaited.

Another recent measurement is that of the radiative decay $B^0_s \to \phi \gamma$, performed by LHCb with $3~\mathrm{fb}^{-1}$ of data at 7 and 8~TeV~\cite{LHCb-PAPER-2019-015}. The chiral structure of the $W$ boson leads to a photon polarisation mostly left-handed in the SM, with a small right-handed component. New physics might significantly alter the contribution of the right-handed component to the total amplitude. LHCb has measured now for the first time direct and mixing-induced $C\!P$ violation in $B^0_s \to \phi\gamma$ decays, with values compatible with the SM, albeit with large uncertainties.

Moving to rare $B$ decays, $B \to \mu^+ \mu^-$ decays are FCNC- and helicity-suppressed, proceeding via $Z$ penguin and $W$ box diagrams. The value of the branching fraction is particularly sensitive to new-physics scalar contributions, such as extra Higgs doublets. CMS and LHCb performed in 2015 a combined fit to their full Run-1 data sets, observing the $B^0_s \to \mu^+\mu^-$ decay for the first time at $6.2\sigma$~\cite{Nature 522 (2015) 68}. In 2016 ATLAS published with Run-1 data~\cite{Eur. Phys. J. C 76 (2016) 513} and in 2017 LHCb performed the first measurement using Run-2 data~\cite{Phys. Rev. Lett. 118 (2017) 191801}. A new measurement by ATLAS, using $26.3~\mathrm{fb}^-1$ at 13~TeV, has been published recently~\cite{arXiv:1812.03017}. By combining Run-1 and Run-2 results, the branching fractions are found to be compatible with the SM at $2.4\sigma$. The analysis of 2016 and 2017 data by CMS is in preparation. In total, 433 $B^0_s$ and 54 $B^0$ candidates are expected with full Run-2 statistics.

The sector of lepton-flavour universality (LFU) tests has received great attention during the last few years. Focusing on $b\to s l^+l^-$ transitions, one measure the ratios $R_K = \mathrm{BR}(B^+ \to K^+\mu^+\mu^-) / \mathrm{BR}(B^+ \to K^+ e^+ e^-)$ and $R_{K^\ast} = \mathrm{BR}(B^0 \to K^{\ast 0}\mu^+\mu^-) / \mathrm{BR}(B^0 \to K^{\ast 0} e^+ e^-)$. The theoretical predictions of such ratios are very clean, hence an observation of non-LFU would be a clear sign of new physics. The measurements are presently at about $3\sigma$ from the SM, complementing a range of other anomalies in $b \to s l^+l^-$ transitions, namely branching fractions in various decays and the angular analysis of $B \to K^\ast \mu\mu$ decays. A new measurement of $R_K$ has been performed by LHCb, adding $2~\mathrm{fb}^{-1}$ of Run-2 data to $3~\mathrm{fb}^{-1}$ of Run-1 data~\cite{arXiv:1903.09252}. The statistics has been doubled with respect to the previous measurement, and the result is $R_K = 0.846^{+0.060+0.016}_{-0.054-0.014}$. In practice, the significance of the discrepancy from the SM is unchanged after the new measurement, as the uncertainty has been reduced but the central value has moved closer to the SM. The outlook is to include also 2017 and 2018 data, for further doubling the statistics, and to add more channels, first to measure $R_{K^\ast}$ with full Run-2 statistics, but also to provide first measurements with $B^0_s$ and $\Lambda_b$ channels.

LFU tests are also performed with semitauonic $B \to D^{(\ast)}\tau\nu$ decays. Here one measures the ratios $R_{D^{(\ast)}} = \mathrm{BR}(B \to D^{(\ast)} \tau \nu / \mathrm{BR}(B \to D^{(\ast)} \mu \nu$, sensitive to new physics at tree level. Measurements of $R_D$ and $R_{D^\ast}$ have been performed by BaBar, Belle and LHCb. The overall average shows a discrepancy from the SM by about $3.8\sigma$. LHCb can also perform measurements with other $b$ hadrons: \emph{e.g.} $B^0_s$, $B_c$ and $\Lambda_b$ decays will help better understand the global picture. A new measurement from Belle has been performed recently, reporting the most precise determination of $R(D)$ and $R(D^\ast)$ to date, and in particular the first $R(D)$ measurement realised using the semileptonic tag. The new results are compatible with the SM at $1.2\sigma$. The $R(D)-R(D^\ast)$ Belle average is now at $2\sigma$ from the SM prediction, and the overall tension with the SM expectation decreases to about $3.1\sigma$.

Concerning the upcoming future, Belle II has concluded successfully the phase-2 pilot run. The basic detector performance is as good as expected, and the nano-beam scheme of the collision point has been tested with encouraging results. The target for phase 3 is to have a full physics run and collect $20~\mathrm{fb}^{-1}$ of luminosity by summer 2019. These are crucial years to demonstrate the capability of the machine to provide the required luminosity while keeping background under control.

\section{Kaon physics}
While waiting for the measurement of $\mathrm{BR}(K^+ \to \pi^+ \nu \nu)$, NA62 performed a measurement using the kaon beam as a source of $\pi^0$s to search for an invisible massive dark photon, $A^\prime$~\cite{arXiv:1903.08767}. No significant excess is detected, using only 1\% of the available statistics collected in 2016-2018. Upper limits are set at 90\% C.L., compatible with fluctuations from the background-only hypothesis. The analysis improves on previous limits over the mass range 60--110 MeV/$c^2$.

Another new measurement in the kaon sector has been performed by NA48 with the $K^\pm \to \pi^\pm \pi^0 e^+ e^-$ decay~\cite{Phys. Lett. B788 (2019) 552}. This is a rare decay proceeding through virtual photon exchange. The first observation has been achieved, and the branching fraction determined to be $(4.237 \pm 0.063\,(\mathrm{stat.}) \pm 0.033\,(\mathrm{syst.}) \pm 0.126\,(\mathrm{ext.}) \times 10^{-6}$.
Several $C\!P$-violating asymmetries and a long-distance $P$-violating asymmetry have been measured and found to be consistent with zero.

\section{Muon magnetic anomaly}
In the SM, the theoretical uncertainty on $a_\mu$ is dominated by hadron vacuum polarisation and hadronic light-by-light scattering. E821 measured $a_\mu$ with a precision of 550 ppb, observing a (statistically dominated) discrepancy from the SM expectation at $3.7\sigma$. E989 is now taking data to reduce the experimental uncertainty to 140 ppb. With such an experimental precision, the reduction of the hadronic theoretical uncertainties will become mandatory. The contribution of hadron vacuum polarisation to $a_\mu$ is calculated through a dispersion relation using the experimental information on the cross-section $\sigma(e^+e^-)\to \mathrm{hadrons}$, as the relevant energy scale is too low for applying perturbative QCD. A reduction of about 20\% in the uncertainty has been achieved since 2013, mostly owing to results from BaBar and VEPP-2000. By contrast, the hadronic light-by-light scattering contribution remains as an open issue, and can be determined with lattice QCD calculations or via further experimental measurements like the proposed MUonE, which aims at measuring the elastic reaction $\mu e\to\mu e$ to constrain $a_\mu^{\mathrm{HLO}}$ from data. Additional physics motivations for light-resonance spectroscopy are the study of hadronic decays of charmonium, via $e^+e^- \to \gamma \psi (\to \mathrm{hadrons})$, the study of other hadronic resonances, searches for dark-photon decays, etc.

\section{Spectroscopy}
\begin{figure}
\begin{minipage}{1\linewidth}
\centerline{\includegraphics[width=0.7\linewidth]{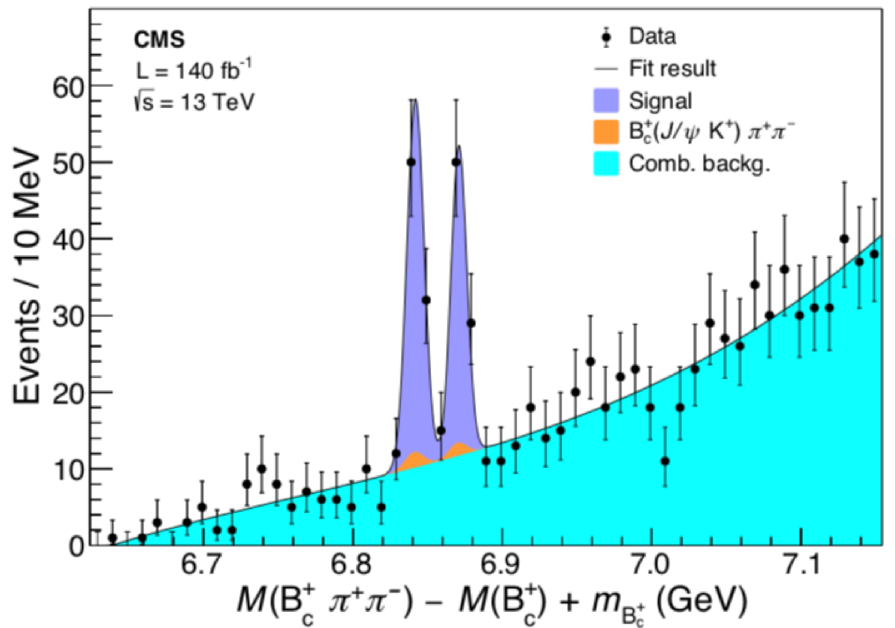}}
\end{minipage}
\caption[]{Invariant $B_c \pi\pi$ mass showing two peaks of excited $B_c$ states~\cite{arXiv:1902.00571}.}
\label{fig:cms_mass}
\end{figure}

A renaissance of QCD in the non-perturbative regime has been taking place during the last decade. A recent measurement with the observation of excited $B_c$ mesons has been performed by CMS, using
$140~\mathrm{fb}^{-1}$ of data at 13~TeV~\cite{arXiv:1902.00571}. Two excited $B_c$ states have been observed in the $B_c \pi\pi$ final state, with $B_c \to J/\psi(\mu\mu)\pi$~(see Fig.~\ref{fig:cms_mass}). The result has been also confirmed by  LHCb, with an analysis on $8.5~\mathrm{fb}^{-1}$ of data at 7, 8 and 13~TeV~\cite{LHCb-PAPER-2019-007}.

\begin{figure}[ht!]
\begin{minipage}{1\linewidth}
\centerline{\includegraphics[width=0.7\linewidth]{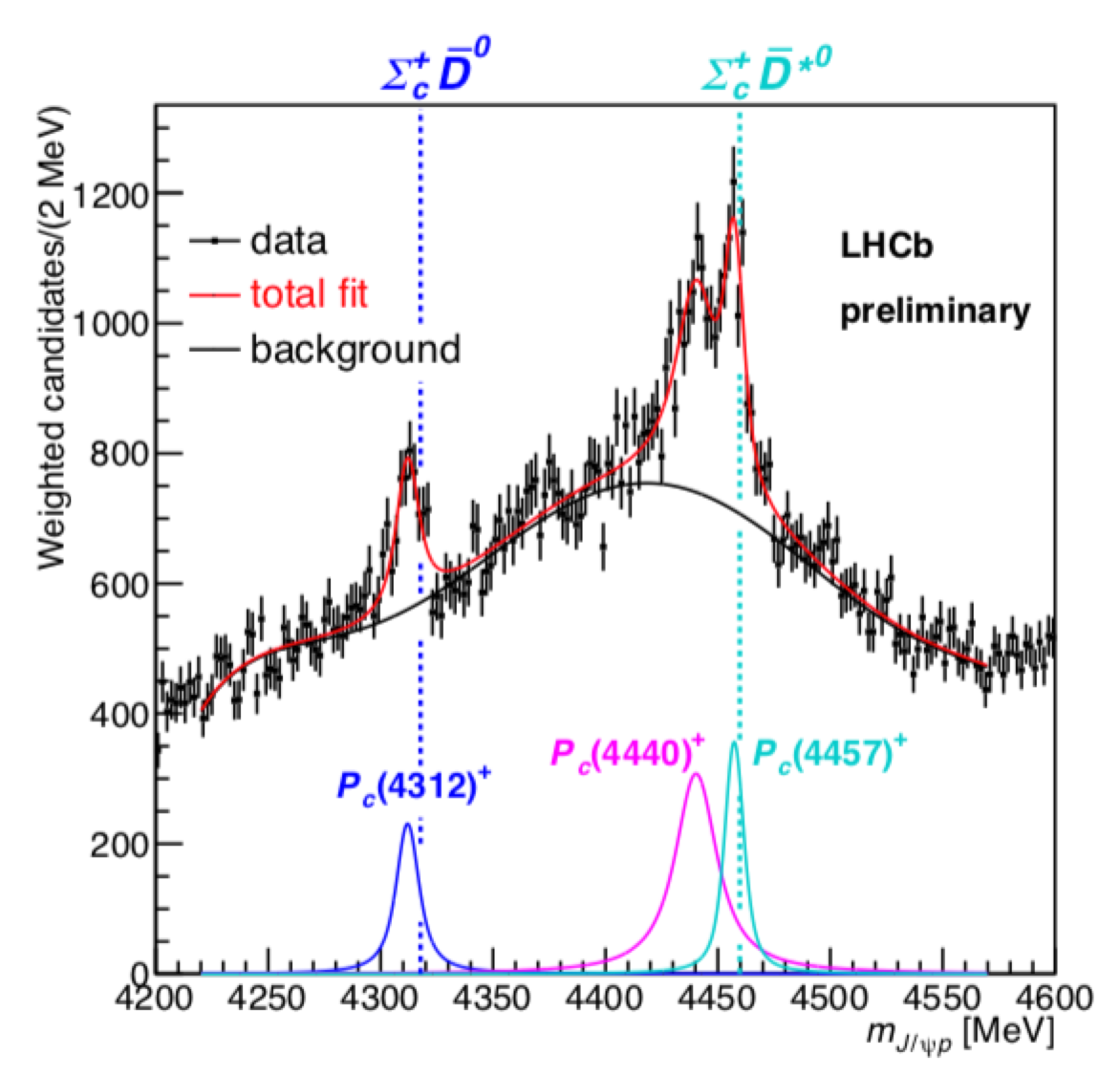}}
\end{minipage}
\caption[]{Invariant $J/\psi p$ mass with the three pentaquark states observed by LHCb~\cite{LHCb-PAPER-2019-014}.}
\label{fig:lhcb_mass}
\end{figure}

One of the most active experiments in spectroscopy measurements with charm quarks is BESIII. BESIII is a dedicated $e^+e^-$ open charm and charmonium factory at BEPC. A plethora of spectroscopy measurements have been performed through the years, notably including tetraquark states. For example, the observation of $X(3872) \to \omega J/\psi$ has been recently reported~\cite{arXiv:1903.04695}. Furthermore, the cross-section measurement of $\gamma X(3872)$ suggests a connection between $X(3872)$ and $Y(4260)$. BESIII is also very active in searching for and studying glueball candidates.

Also LHCb is extremely active, with novel spectroscopy measurements being performed at an impressive rate. LHCb has observed recently a new state, likely to be $\psi(1^3D_3)$, \emph{i.e.} achieving the first observation of a spin-3 charmonium state~\cite{LHCb-PAPER-2019-005}. LHCb has also got relevant news on pentaquarks. First pentaquarks were observed by LHCb four years ago using $\Lambda_b \to J/\psi pK$ decays as a proxy. Two charged states were determined: one narrow, dubbed $P_c(4450)$, and one broader, dubbed $P_c(4380)$, both decaying into $J/\psi p$. The measurement triggered great theoretical interest to understand the nature of the new resonances, \emph{i.e.} whether they are tightly bound or molecular states. LHCb has now updated the same pentaquark analysis using $9~\mathrm{fb}^{-1}$ of data at 7, 8 and 13~TeV~\cite{LHCb-PAPER-2019-014}. The overall statistics has increased by a factor nine with respect to the the Run-1 analysis. Only a narrow bump-hunting analysis with empirical background shape has been performed so far. As shown in Fig.~\ref{fig:lhcb_mass}, the previously found $P_c(4450)$ reveals a finer structure with two close peaks, and a new peak is found at 4312~MeV.
This constitutes an important novel input to shed light into the nature of pentaquarks. A full amplitude analysis is ongoing, especially needed to confirm the $P_c(4380)$ state.

\section{Conclusions}
Despite the formidable attempts performed so far by high-energy physics experiments, the Standard Model of particle physics still proves to be unbreakable. In the current state of fundamental physics, it is of paramount importance to maintain and further develop a diversified programme, ranging from the high-energy frontier to the intensity frontier, from astrophysics and cosmology to dark matter detection. We know that the Standard Model will capitulate in the end, the only question is when and how. Meanwhile, long live Rencontres de Moriond!

\section*{Acknowledgments}
I wish to express my cordial thanks to the organisers of the conference for the kind invitation and for the pleasant environment.

\end{document}